\documentclass[preprint]{aastex}

\newcommand{\co}{$^{12}$CO}
\newcommand{\coratio}{^{12}{\rm C}/^{13}{\rm C}}   
\newcommand{\ocen}{$\omega$ Cen\ }         
\newcommand{\omegacen}{$\omega$ Centauri}  

\slugcomment{}

\shorttitle{Carbon isotopes in $\omega$ Cen}
\shortauthors{Smith et al.}

\begin{document}

\title{Carbon Isotopic Abundances in the Red Giants of $\omega$ Centauri
(NGC 5139)}

\author{Verne V. Smith}
\affil{Department of Physics, University of Texas at El Paso, El Paso,
TX 79968}

\author{Donald M. Terndrup\altaffilmark{1}}
\affil{Department of Astronomy, The Ohio State University, Columbus,
OH 43210}

\and

\author{Nicholas B. Suntzeff}
\affil{Cerro Tololo Inter-American Observatory, National
Optical Astronomy Observatories, Casilla 603, La Serena, Chile}

\altaffiltext{1}{Visiting Astronomer, Cerro Tololo Inter-American
Observatory, National Optical Astronomy Observatories, which are
operated by the Association of Universities for Research in 
Astronomy, Inc., under cooperative agreement with the National
Science Foundation.}

\begin{abstract}

Carbon-12 and carbon-13 abundances have been measured in eleven
bright giant members of the globular cluster \omegacen\ via observations 
of the first-overtone CO bands near 2.3 $\mu$m.  The stars in this 
sample were selected to span a substantial fraction of the range of iron 
abundances found in this cluster.  In addition, the sample covers a 
range of [O/Fe], [Na/Fe] and [Al/Fe] abundance ratios derived 
in previous studies.  In all $\omega$ Cen giants 
the $\coratio$ abundance ratio is found to be quite low,
indicating deep mixing in these red giants.  The mean value
for the entire sample is $\langle\coratio\rangle = 4.3 \pm 0.4$
($\sigma = 1.3$),
with nine stars equal, within the errors, to the equilibrium
ratio $\coratio = 3.5$ and two stars having slightly higher
values.  There is no correlation between the $\coratio$ and the abundance 
of iron.  In addition, no correlation of $\coratio$ with [$^{12}$C/Fe]
is found (all giants are deeply mixed), although the derived abundances
of [$^{12}$C/Fe] show a positive correlation with [O/Fe], and an
anticorrelation with [Na/Fe] (with the oxygen and sodium abundances taken
from previous studies in the literature).  A comparison of the isotopic
carbon ratios in $\omega$ Cen with those from other globular clusters
(M4, M71, NGC6752, and 47 Tuc), and with literature oxygen abundances,
may reveal a slight trend of decreasing $\coratio$ ratios with decreasing
[O/Fe] in the entire globular cluster sample of red giants.  A comparison
between $^{12}$C/$^{13}$C and both [Na/Fe] and [Al/Fe], however, reveals
no trend.
\end{abstract}

\keywords{globular clusters: individual ($\omega$ Centauri)--nuclear
reactions, nucleosynthesis, abundances--stars; abundances--stars:
late-type--stars: Population II}

\section{Introduction}

In the last two decades, it has become clear that evolved
giants in metal-poor globular clusters often exhibit a wide
range in the abundances of light elements 
(Paltoglou \& Norris 1995; Norris \& Da Costa 1995b; Pilachowski,
Sneden, \& Kraft 1996; Shetrone 1996a, b; Kraft et al. 1997;
Sneden et al. 1997; Gonzalez \& Wallerstein 1998; Ivans et al. 1999;
Smith et al. 2000).
These variations are generally agreed to arise from
proton capture chains that convert C and O into N,
Ne into Na, and Mg to Al in the hydrogen-burning layers
of red giants.  There has been considerable discussion,
however, as to the origin of these abundance patterns
(e.g., Cottrell \& Da Costa 1981; Kraft 1994; Norris \& 
Da Costa 1995a; Sneden et al. 1997).  In the
``primordial'' scenario, the abundance patterns reflect
nucleosynthesis in a prior generation of massive stars;
support for this comes from variations in the C, N, O, Na, Mg, and 
Al abundances among both main-sequence and turn-off stars in various 
globular clusters, as found by Bell, Hesser, \& Cannon (1983),
Briley, Hesser, \& Bell (1991), Suntzeff \& Smith (1991), 
Briley et al. (1996), Cannon et al. (1998) or Gratton et al. (2001).  
In contrast,
the ``evolutionary'' scenario envisions that the products
of internal proton capture are brought to the stellar
surface through mixing in the giants now observed in the clusters.  
Evidence in favor of this comes
from abundance patterns that depend on the evolutionary
state of the stars in a cluster.  For example, several
systems show a decline in the overall carbon
abundance with increasing stellar luminosity and
a corresponding rise in the abundance of nitrogen,
a sharp decline in the ratio of $\coratio$ from
the presumed primordial value near 90 (i.e., the solar
value) to values as low as the nuclear equilibrium
value of 3.5.  This equilibrium ratio of $^{12}$C/$^{13}$C= 3.5,
which is set largely by the ratio of reaction rates
$^{12}$C(p,$\gamma$)$^{13}$N($\beta$$^{+}$,$\nu$)$^{13}$C and
$^{13}$C(p,$\gamma$)$^{14}$N, is insensitive to temperature and
is a well-determined astrophysical limit (see the review by Wallerstein
et al. 1997).  So-called ``standard'' stellar models evolving up the
first ascent of the red giant branch, e.g., Iben (1964) or
Charbonnel (1994), predict values to go only as low as
$^{12}$C/$^{13}$C$\sim$ 20: the differences between the observed (low)
isotopic carbon ratios and the higher values predicted by the standard
models of stellar evolution has been a longstanding piece of evidence
used to invoke extra mixing (or deeper mixing) in red giants. 
Quite possibly, in the globular cluster stars, both primordial 
and evolutionary chemical
evolution occurs, whereby deep mixing 
in the current giants is superimposed on preexisting abundance
patterns (Briley et al. 1994; Ivans et al. 1999).

Mixing signatures are seen also in metal-poor
field giants (Gratton et al. 2000, and references therein),
with some interesting differences found 
between the field stars and globular cluster giants
of the same overall metallicity.  The O--Na and Mg--Al
anticorrelations found in some of the globular clusters
are not observed in any of the field red giants.  One effect
noted in the field giants by Gratton et al. (2000), and also
predicted by current non-standard stellar models (Charbonnel 1994;
Wasserburg, Boothroyd, \& Sackmann 1995; Charbonnel, Brown, 
\& Wallerstein 1998), is that the
efficacy of first giant branch mixing will increase as the
metallicity decreases. 
The globular cluster $\omega$ Centauri is a site where the
various mixing trends (as a function of [Fe/H] and as a
function of Na and Al variations) can be studied in one object.

The most massive globular
in the Milky Way, $\omega$ Cen has a wide spread of [Fe/H]\footnote{
Throughout this paper, we will use the normal spectroscopic
notation that ${\rm [A/B]} \equiv \log_{10}(N_A / N_B)_{\rm star}
- \log_{10}(N_A / N_B)_\odot$, for elements A and B.} 
(Suntzeff \& Kraft 1996; Norris, Freeman, \& Mighell 1996, 
and references therein), 
unique among globulars, with the possible exception of M22 (which may
show an abundance spread, but much smaller than in $\omega$ Cen).  In 
$\omega$ Cen the distribution of [Fe/H] has a floor near --2.0 to --1.8,
presumably representing the initial metallicity of the gas out
of which the cluster was formed, and shows an extended
tail to higher metal abundances.   The
abundances of most elements increase with [Fe/H] and
exhibit relatively small scatter, with the exception of
O, Na, Al and some other light elements (Brown \& Wallerstein 1993; 
Norris \& Da Costa 1995b; Smith et al. 2000).  As in other 
globular clusters, the giants of $\omega$ Cen have
[Na/Fe] and [Al/Fe] abundances which are positively
correlated with each other, but anticorrelated with
[O/Fe] in a way that is most easily (but not
definitively!) explained as a product of deep mixing.
The abundance patterns of elements heavier than Fe
suggest nucleosynthesis from AGB stars between 1 and 3$M_\odot$,
but whether this occurred during formation of the
cluster that involved self-enrichment, or from 
mergers of fragments with different chemical abundances,
is not yet determined.

To date, only a few measurements of the $\coratio$ isotopic ratio
exist for $\omega$ Cen giants (3 from Brown \& Wallerstein 1993).
Their 3 values were all quite low, with $^{12}$C/$^{13}$C= 6, 4, and 4;
however, the metallicities sampled by these 3 giants was limited
([Fe/H]= -1.36, -1.34, and -1.26).
The goal of this work is to derive additional values of $\coratio$
in a larger sample of $\omega$ Cen giants, in order to investigate if 
there are any trends with [Fe/H], or if this somewhat unique globular
cluster differs in its $\coratio$ ratios when compared to other clusters. 

\section{Observations and Data Reduction}

The program stars were selected from the high-resolution
study of Norris \& Da Costa (1995b), and were chosen to cover a wide
range of [Fe/H] at magnitudes where spectroscopy at $2.35 \mu$m 
was practical with a 4m telescope.  Most of these stars also have
other metallicity determinations from various studies
employing low- to high-resolution spectra.
Table 1 summarizes the available data for our
targets.  The first column of that table lists the
name of each star from Woolley (1966).  The second and third
columns list photometry from several sources as assembled
by Suntzeff \& Kraft (1996);  for three stars not tabulated in that
paper (ROA 150, 155, and 161), we took the photometry
from Cannon \& Stobie (1973) and applied a reddening correction of
$E(B - V) = 0.11$ (Butler, Dickens, \& Epps 1978).  Infrared photometry
from Persson et al. (1980) follows in column 4, while the remaining
columns show determinations of [Fe/H] from papers listed
in the table, all of which are evidently on the same
metallicity scale (an exception are the values of Francois, Spite, \& 
Spite (1988), not tabulated here, which are about 0.2 dex lower than 
in these other studies).  We will use the Norris \& Da Costa (1995b) 
values of [Fe/H] throughout the remainder of this paper.

In Figure 1 we present a color-magnitude
diagram of $\omega$ Cen. The small
points show cluster members ($\geq 50\%$ membership
probability) from van Leeuwen et al. (2000), where
the photometry comes from sources referenced in that
paper.  The large symbols show photometry for
our targets.  Most of our stars are brighter than the
magnitudes where the RGB and AGB are clearly separated,
but most should be on the RGB since the evolutionary
lifetimes there are much longer than on the AGB.  

An additional selection criteria was to observe stars
that spanned the range of Na and Al abundances, which
in $\omega$ Cen and other globular clusters are often anticorrelated
with oxygen.  

Spectra of the first-overtone bands of CO near
2.3 $\mu$m were obtained on June 2--4, 1996, and in
a second run on June 10-12,
1998 with the
Blanco 4m Telescope and the CTIO infrared spectrometer
(DePoy et al. 1990).
The 1996 run used an engineering-grade $256 \times 256$
InSb detector from SBRC which had many defective
pixels and an appreciable dark current.  The 1998 run
employed a similar detector of higher quality, and
also was obtained with a tip-tilt secondary. The length of the
slit was 16$\arcsec$, and the slit width was set to
0.7$\arcsec$. The
pixel size for both runs was 24 $\mu$m for a resolution
of $\lambda / \Delta\lambda \approx 9900$ pixel$^{-1}$. 
The effective resolution was obtained from 
observations of telluric emission lines, which had
an average FWHM of 2.7 pixels.
A $K$ filter was used for order separation.

The 1996 observations were obtained in three overlapping
wavelength intervals, which are shown in Table 2.  In
1998, we obtained spectra at two wavelength settings.
These covered the (2--0), (3--1), and (4--2) bands of
$^{12}$CO and the (3--1) band of $^{13}$CO and also
included features of Ca and Mg (Kleinman \& Hall 1986);  the latter
are, however, very weak in our spectra of the \ocen targets 
and will not be discussed further.  Figure 2 shows two sample
calibrated spectra and the identification
of the more prominent features.  These two giants have similar
effective temperatures and metallicities, but note that ROA 155
has noticably stronger CO bands than ROA 139.  Norris \& 
Da Costa (1995b) classify ROA 139 as CN-strong and ROA 155 as
CN-weak; the difference between the CN bandstrengths represents
differences in the nitrogen abundance (Norris \& Da Costa find
[N/Fe]= +1.05 for ROA 139 and +0.40 for ROA 155).  The differing
CO bandstrengths observable in Figure 2 fit in with the C--N
anticorrelation expected from differing degrees of CN-cycle mixing. 

The observing procedure in both years
was to obtain repeated spectra with the star placed at 
various locations along the slit.  Exposure times
ranged from 30 to 60 seconds at each position, and the
motion of the star along the slit between exposures
was sufficient to completely separate the spectra on
the detector. 
In both runs, observations of a hot
star (HD~116717, spectral type A0V) were interleaved 
with observations of the $\omega$ Cen targets.  
The observing procedure was identical
to that used for the program stars, except that since HD 116717
was brighter, the exposure times were shorter and fewer
positions along the slit were needed in order to achieve the
same, or higher, S/N than in the $\omega$ Cen stars.  Spectra were
obtained for several stars at one wavelength interval, then
repeated at the other wavelength settings.

Data reduction was performed using scripts written for
the IRAF\footnote{The IRAF software is distributed by the
National Optical Astronomy Observatories under contract with
the National Science Foundation} and VISTA packages.
The images were first corrected for the nonlinear response
of the detectors.  The correction multiplied the raw
counts by a quadratic polynomial in intensity.  The correction
was almost always less than 0.5\%, but was 2--3\% at the
center of the spectrum for the brightest stars.  Next, the
several images for each star were combined via a median,
which removed the individual spectra.  This was subtracted
from the images, thereby correcting for the sky and the
dark current.  Finally, the images were divided by a flat
field, which was obtained by combining many images of a
white spot on the telescope dome, obtained in two passes 
with the dome lights on and off then subtracted.

The resulting spectra were then traced, extracted in
an aperture of width twice the FWHM of the spectra,
and co-added after multiplicatively scaling to the
average mean exposure.  The co-adding process computed
an average with rejection of pixels that were $> 4\sigma$
away from the average.  The deviation about the mean
gave a measure of the S/N per spectrum, which was typically 75
to 120 for the $\omega$ Cen spectra.

The final step was to correct for the many telluric
absorption lines in the spectra using the observations of
the spectra of the hot star.  When hot star spectra
were taken both before and after one of the $\omega$ Cen
targets, we interpolated the spectra linearly in time 
to the midpoint of the target star's observation.  If
spectra were not available that bracketed an observation
in \ocen, we always had one adjacent to the observation.  We
measured the shift in the wavelength direction (in
pixels) between the interpolated spectrum and the combined
target spectrum to ensure that they were aligned correctly.
The resulting pixel shifts were almost always less than
0.3 pixel.  Then the object spectrum was divided by
the interpolated spectrum.  

Because there are few lamp arc lines in the
spectral intervals we observed, the
wavelength calibration was achieved in several steps.  We
first computed the linear wavelength solution
for the bluest observed wavelength interval in the 1996
spectra, in which there are several OH airglow lines that have
wavelengths compiled by Oliva \& Origlia (1992).  The dispersion
($\mu$m pixel$^{-1}$) from this solution was assumed to
be the same for the other two wavelength intervals.
We determined the zero point for the central wavelength
interval by computing the average pixel shift between
the two spectrum sections via cross correlation. 
Finally, the zero point for the red spectral interval
was adjusted to match the 
the 3--1, 2--0 and 4--2 bandheads of \co, where
the wavelengths of these were taken from Kleinman \& Hall (1986).  
A similar procedure was followed for the 1998 data,
using the airglow lines in the blue spectra to set
the dispersion for both wavelength intervals, then
adjusting the zero point for the red spectra to match
the \co\ bandheads.  The r.m.s.\ residuals in the solutions using
either the telluric lines or the CO bandheads were
always less than 0.4 pixels.

\section{The Carbon Isotopic Abundance Analysis}

An abundance analysis of the carbon isotopes, using the CO features,
was carried out for the program stars using the most recent version
of the LTE spectrum synthesis code MOOG (Sneden 1973).  Model atmospheres
were computed using a version of the MARCS code described in
Gustafsson et al. (1975), and are plane parallel, hydrostatic equilibrium
models.  For the $\omega$ Cen giants spanning the range in temperature,
gravity, and metallicity found here, the use of MARCS--type models is
appropriate, as discussed in a number of abundance studies, e.g., Smith
et al. (2000), Ramirez et al. (2001), or Cunha et al. (2002).  In
addition, an analysis by Hinkle \& Lambert (1975) of the formation
of the CO vibration-rotation (V--R) lines in the types of red-giant
atmospheres studied here finds that these lines form to a high degree
in LTE. 

As mentioned in the Introduction, all of the $\omega$ Cen program
stars observed here were analyzed by Norris \& Da Costa (1995b), thus
detailed abundances from a number of elements are already available.
The stellar parameters of effective temperature (T$_{\rm eff}$), surface
gravity (log g), microturbulence ($\xi$), and metallicity (taken as
[Fe/H]) were derived by a combination of photometric and spectroscopic
indicators, as discussed by Norris \& Da Costa (1995b).  A comparison
of those parameters, with those derived 
independently by Smith et al. (2000)
for a subset of the Norris \& Da Costa (1995b) sample, is 
described in both Smith et al. (2000) and Cunha et al. (2002).  No
unexpected systematic errors or uncertainties are found and 
T$_{\rm eff}$, log g, $\xi$, and [Fe/H], as derived by Norris \&
Da Costa (1995b) are adopted here.  Estimates of uncertainties in
the stellar parameters, as discussed in the various abundance analysis
papers mentioned above, are about $\pm$100K in T$_{\rm eff}$,
$\pm$0.2 dex in log g, $\pm$0.2 km s$^{-1}$ in $\xi$, and $\pm$0.05 dex
in [Fe/H].  The adopted stellar parameters are shown in the beginning
columns of Table 3. 

Spectral--line lists covering the regions observed were constructed
using the atomic linelist from the Kurucz \& Bell (1995) dataset
and CO lines from Goorvitch (1994).  Lines due to CN in this spectral
region were also included and were kindly provided by B. Plez
(2001, provate communication).  For giants with the temperatures and
metallicities of our program stars, the CN lines are a minor contribution,
even for very large nitrogen enhancements.  In all of the program stars
the C/O abundance ratio is less than 1.0 and CO
formation is controlled primarily by the carbon abundance.  Although
the derived carbon abundance from CO depends slighty on the oxygen
abundance, this dependence is not large for the temperature regime
covered here and oxygen abundances are 
available from Norris \& Da Costa (1995b), who used the [O I] 
$\lambda$6300\AA\ line.  With a given O abundance and model atmosphere,
the CO bands were synthesized as a function of carbon abundance.  
Synthetic spectra were calculated with wavelength increments of 0.01\AA\
(or $\lambda$/$\Delta$$\lambda$= 2.3$\times$10$^{6}$) and smoothed with
a Gaussian broadening function of 6.4\AA, corresponding to the measured
widths of intrinsically narrow telluric emission lines.  Typical
internal stellar broadening mechanisms in these giants (thermal,
microturbulence, and macroturbulence) are $\sim$ 0.8-1.0\AA\ at this
wavelength, so instrumental broadening is by far the dominant term.
The Gaussian smoothing function was found to provide excellent fits to
the observed spectra.
The
$^{12}$C abundance was derived first, giving most weight to the regions
covering the $^{12}$CO (3--1) and (4--2) bandheads, as these are free
from $^{13}$CO contamination and were found to be most sensitive to the
carbon-12 abundance.  A sample comparison and fit between observed and
synthetic spectra is shown in Figure 3 for the star ROA155 ([Fe/H]= -1.64).
The top panel shows the $^{12}$CO (3--1) bandhead for three different
$^{12}$C abundances.  Straightforward least-squares residuals were used
to select the best--fit carbon abundance, and this abundance was rounded to
the nearest 0.1 dex.  Residuals were computed for regions covering
$\sim$23225-23260\AA\ for $^{12}$CO (3--1) and 23515-23550\AA\ for 
$^{12}$CO (4--2).  Each
region contains about 17 data points, so the minimum sum of squared
residuals was not affected significantly by a dew deviant points.
The bottom panel illustrates the fit for the
$^{13}$CO (2--0) bandhead in this star, using the already derived $^{12}$C
abundance; the $^{13}$C abundance is parameterized by the 
$^{12}$C/$^{13}$C ratio and synthetic spectra computed for three different
isotopic ratios are shown, with $^{12}$C/$^{13}$C= 4.0 being the best
fit.  The interval for calculating residuals was
$\sim$23447-23480\AA\, containing 16 or 17 data points: the
resulting minima were not influenced significantly by the occasional
noisy point.  Isotopic ratios were rounded to the nearest 0.5.   

The primary uncertainties in the derived carbon isotopic abundances arise
from uncertainties in the input stellar parameters T$_{\rm eff}$, log g,
and microturbulence (the model metallicity has negligible effect on the
derived abundances, unless the change in metallicity is substantial, i.e.
factors of 10).  Tests were done for parameter changes of 100K in
effective temperature, 0.2 dex in log g, and 0.2 km s$^{-1}$ in
$\xi$, with the following results for the change in carbon-12
abundances: $\Delta$T$_{\rm eff}$= +100K yields $\Delta$C= +0.05 dex,
$\Delta$(log g)= +0.2 dex yields $\Delta$C= +0.05 dex, and
$\Delta$$\xi$= +0.2 km s$^{-1}$ yields $\Delta$C= -0.04 dex.  An additional
uncertainty arises from the S/N of the spectra, and based on the
depth and width of residual minima as a function of carbon abundance,
a typical 1$\sigma$ uncertainty is $\pm$0.04 dex in $\Delta$C.  Compared
to all of these errors, broadening adds an insignificant uncertainty
as the total integrated absorption is conserved and red giants with
these temperatures and abundances hasve flux points close to the
continuum blueward of the $^{12}$CO (3-1) bandhead.  Adding
these abundance errors in quadrature provides an abundance uncertainty
of $\pm$0.09 dex for carbon.  The isotopic ratios are very insensitive
to the model parameters, but a conservative estimate based on testing
various fits indicates that the ratio itself is uncertain by $\pm$1.0
in $^{12}$C/$^{13}$C.  For all program stars here, 
Norris \& Da Costa (1995b) derived $^{12}$C abundances from the violet 
CH bands, and a comparison is shown (with both our estimated errors and
their estimated errors) in Figure 4.  The comparison is quite good,
with the mean difference (Us -- Norris \& Da Costa) being +0.14 $\pm$0.24
dex.  
One star, ROA 139, is the only giant to show a 
discrepancy (we find [C/Fe]= -1.13 while Norris \& Da Costa
derive -0.50).  It is worth noting that this star has the most extreme
nitrogen abundance found by Norris \& Da Costa, with [N/Fe]=+1.05.
In a plot of [N/Fe] versus [C/Fe], assembled from the Norris \&
Da Costa abundances, ROA 139 stands out as falling well above their
mean relation (i.e., quite a large [N/Fe] ratio for their value of
[C/Fe]).  Simply extrapolating their [N/Fe]--[C/Fe] relation to
a large value of [N/Fe] would suggest that ROA 139 should have a
very low [C/Fe] ratio ($\le$ -1.0), as we find.  It may be that
at very low carbon abundances, the violet CH-band becomes quite
weak.  

The Norris \& Da Costa estimated uncertainties of $\pm$0.21 dex,
coupled with our uncertainties of $\pm$0.10 dex (in [C/Fe], assuming
an uncertainty of 0.05 dex in Fe), yield an expected scatter of
0.23 dex, close to the observed value of 0.24 dex.  The small offset
of +0.14 dex in the absolute abundance scale probably arises from
small differences in the gf-value scales between CH and CO, as well as
a possible small difference in the adopted solar carbon abundance.
A recent value for the solar carbon abundance is by Holweger (2001),
who finds log $\epsilon$(C)= 8.59.  This abundance can be checked
from a solar analysis using the same CO lines as those used in our
$\omega$ Cen analysis.  We generated a MARCS solar model (T$_{\rm eff}$=
5777K, log g= 4.438) with a microturbulent velocity of 1.0 km s$^{-1}$
(this value of $\xi$ yields our preferred iron abundance of
log $\epsilon$(Fe)=7.50) from an analysis of both Fe I and Fe II lines
with well-defined laboratory oscillator strengths, and is the same
solar Fe abundance adopted by Norris \& Da Costa (1995b), whose
$\omega$ Cen Fe-abundance scale is used here).  The solar flux spectrum
used is that from Wallace \& Hinkle (1996), which has a spectral resolution
of $\lambda$/$\Delta$$\lambda$= 261,000.  Syntheses of the same regions
used in the $\omega$ Cen analysis (23225-23260\AA\ and 23515-23550\AA)
were examined and the best solar $^{12}$C abundance derived from the
$^{12}$CO lines (using a MARCS model) is log$\epsilon$(C)= 8.55: this is
close enough to the Holweger (2001) value that we will adopt 8.59 as
the solar $^{12}$C abundance.
the two respective solar Fe abundances are the same, with 
log $\epsilon$(Fe)= 7.50).  This comparison indicates that the carbon
isotopic abundances presented here do not suffer from significant
systematic effects.  The carbon-12 abundances, as well as the
$^{12}$C/$^{13}$C ratios are listed in Table 3.  The [Fe/H] values from
Norris \& Da Costa (1995b) are used to also tabulate values of
[$^{12}$C/Fe].  

\section{Results and Discussion}

An initial investigation into the isotopic carbon abundance ratios
in these $\omega$ Cen giants is illustrated in Figure 5.  In the
top panel, $^{12}$C/$^{13}$C ratios are plotted as a function of
[Fe/H] for various samples of metal-poor giants.  A number of 
globular clusters have been analyzed for isotopic carbon ratios and
we include in this discussion the results from Smith \& Suntzeff (1989) 
and Suntzeff \& Smith (1991) for M4 (39 giants), Suntzeff \&
Smith (1991) for NGC6752 (12 giants), Bell, Briley, \& Smith (1991)
for 47 Tuc (4 giants), and Briley et al. (1997) for M71 (10 giants).
In addition, samples of field stars have been analyzed and new results,
as well as a summary of earlier work, have been presented most recently
by Keller, Pilachowski, \& Sneden (2001): these results are also
included in the comparisons.  Concentrating first on the 
$\omega$ Cen giants, there is no significant trend
of $^{12}$C/$^{13}$C with [Fe/H], and all member giants show
evidence of deep mixing.   Two of the more metal-poor $\omega$
Cen giants have slightly larger ratios (6 for ROA 252 and
7 for ROA 58), however, the overall carbon abundances in these
particular stars are quite low and the CO bands weaker.  The
$^{13}$CO bands in these two giants are weaker than in the
other program stars, thus the uncertainties are probably
somewhat larger: before a trend could be claimed, a larger sample
of $\omega$ Cen giants would need to be studied.  With the
current results, no real trend with [Fe/H] is found for the
$\omega$ Cen members.  

In general,
the $\omega$ Cen results look very similar
to what is found for the other globular clusters, with the following 
note: the two fairly metal-rich clusters, M71 and 47 Tuc, appear to
have substantial fractions of their members with both slightly high
($^{12}$C/$^{13}$C $\sim$ 7-8) and low ($^{12}$C/$^{13}$C $\sim$ 4)
ratios.  Briley et al. (1997) have shown that these ratios are
anticorrelated with CN band strengths (the CN-strong giants have lower
$^{12}$C/$^{13}$C ratios and the CN-weak giants have higher ratios).   
This dichotomy is not seen in the somewhat more metal-poor clusters
M4 and NGC 6752 (although this cluster does have 1 giant in the sample
of 12 with a high ratio and it is one of the CN-weak stars).  No
measurable effect in the C-isotopic ratios as a function of CN strength
is found for $\omega$ Cen: $\coratio = 4.9 \pm 0.8$ 
and $3.5 \pm 0.5$ (m.e.) for the CN-weak and CN-strong stars,
respectively: a difference of only $1.5\sigma$.   

An additional observation that should be noted
about the top panel of Figure 5 is that the field-star sample contains
a number of quite metal-poor giants with $^{12}$C/$^{13}$C ratios of
about 8-10, somewhat higher than most of the globular-cluster giants.
This point is addressed in the bottom panel of Figure 5, where
$^{12}$C/$^{13}$C is plotted versus absolute visual magnitude.  The
field giants with the slightly larger C-isotopic ratios tend to be
somewhat less luminous than the globular-cluster sample.  The luminosity 
dependence of $^{12}$C/$^{13}$C is discussed by Keller et al. (2001)
and the isotopic ratios in the more luminous field giants overlap
perfectly with the globular-cluster giants.  Again, the $\omega$ Cen
member ROA 58 (with $^{12}$C/$^{13}$C= 7.0) stands out somewhat in
this plot; however, as discussed above, this giant has a low carbon
abundance and the CO bands are weak: better spectra would need to be
obtained, or more $\omega$ Cen giants observed, in order to decide
whether this result for ROA 58 is real or significant.  The conclusion 
from this study is that for these rather luminous $\omega$ Cen giants,
with M$_{\rm V}$$\le$ -1.8, there is no measurable change in
$^{12}$C/$^{13}$C versus [Fe/H].

As discussed in Section 3, the absolute carbon abundances 
can also be derived fairly accurately for these giants from the
CO bands.  In Figure 6, a number of isotopic and elemental abundance
ratios are plotted versus [$^{12}$C/Fe] (hereafter [C/Fe]): 
the carbon-12 abundance is predicted to decrease as the envelope
fraction of CN-processed material increases.  In the mixing scenario
a lower [C/Fe] corresponds to deeper mixing, while in the primordial
scenario it results from incresed pollution from material processed
through presumed progenitor stars.  The top left panel in
Figure 6 shows $^{12}$C/$^{13}$C versus [C/Fe]
for $\omega$ Cen and a number of other samples.  Concentrating
on the $\omega$ Cen members for the moment, no trend is
found, with all $\omega$ Cen giants having low $^{12}$C/$^{13}$C ratios,
indicative of envelope material heavily exposed to the CN-cycle, and 
$^{12}$C to Fe ratios spanning about a factor of 10.  Correlations
between $^{12}$C, O, and Na are explored in the bottom left and top
right panels of Figure 6.  In these panels there is clear correlation
between [O/Fe] versus [C/Fe] and an anti-correlation between
[Na/Fe] versus [C/Fe] (the Na and
O abundances for $\omega$ Cen are from Norris \& Da Costa 1995b). 
As discussed in the Introduction, these are the trends expected
for material exposed to H-burning. 
Anticorrelations between [Na/Fe]
and [O/Fe], are found in many globular clusters, e.g., Langer, Hoffman,
\& Sneden (1993), Kraft et al. (1993), Pilachowski et al. (1996), or
Kraft et al. (1998).  One interpretation of these Na--O anticorrelations
is deep mixing in the currently observed giants, although Briley et
al (1996) find Na-CN anticorrelations in 47 Tuc turnoff stars, while
Cannon et al. (1998) find C-N anticorrelations in 47 Tuc main-sequence
stars, and 
Gratton et al. (2001) have found Na--O anticorrelations in
main-sequence members of NGC6752.  The ultimate origins of these
various abundance correlations and anticorrelations observed in a
number of globular clusters are still not understood completely.

Other globular clusters are plotted in Figure 6 as comparisons to
$\omega$ Cen.  In the top left panel, results are available for
five other clusters, as well as field stars.  Omega Cen
does not obviously stand out relative to the other globular clusters.
In the bottom left panel of Figure 6 results for the clusters M4
and M71 are compared to those from $\omega$ Cen; the [O/Fe] values for 
M4 are from Ivans et al. (1999) and for M71 from Briley et al. (1997).
Oxygen abundances are not available for the samples we are using
as comparisons for NGC6752 and 47 Tuc.  All three globular clusters
plotted show a large degree of overlap in the trend of oxygen versus
carbon; however, $\omega$ Cen displays a larger range in values of both
[C/Fe] and [O/Fe].  
Along a similar line, the top right panel of Figure 6 shows [Na/Fe]
versus [C/Fe] for the $\omega$ Cen stars, along with the M4 sample
(with the [Na/Fe] coming from Ivans et al. 1999).  An anticorrelation
of Na and C is found in both clusters and the trends are, within
the uncertainties, identical.  Again, though, $\omega$ Cen displays a
larger range in values of [Na/Fe] relative to M4. 
Greater differences in [C/Fe], [O/Fe], and [Na/Fe] could indicate
either deeper mixing (in the mixing scenario) or, on the other hand, 
larger amounts of pollution from material processed through H-burning
(in the primordial scenario).  In a recent study of M71, Ramirez \&
Cohen (2002) have summarized [Na/Fe] and [O/Fe] trends for 11
globular clusters (their Figure 12) and these different clusters
display a range in their respective values of both [Na/Fe] and [O/Fe].  
If this range
is quantified simply as the difference between the largest and smallest
values, a weak trend is found between a larger range in [Na/Fe] and
[O/Fe] with decreasing [Fe/H].  It is not clear how $\omega$ Cen
fits into this possible trend, as these other clusters have no spread
in [Fe/H], whereas the spread in $\omega$ Cen exceeds a factor of 20;
however the range in O and Na to Fe ratios displayed by $\omega$ Cen
members is as large as any found in the other globular clusters, with 
M13 and M92 showing the largest ranges. 

A somewhat different type of chemical behavior is probed in the
bottom-right panel of Figure 6, where a slow neutron-capture (s-process)
element (La) is plotted versus [C/Fe] for both $\omega$ Cen and M4.
Omega Cen is peculiar in that its member stars display a large increase,
in proportion to Fe, of s-process elements as [Fe/H] increases: such
behavior is not observed in any other stellar population.  The
increase in s-process abundances originates from thermally-pulsing
asymptotic giant branch (TP-AGB) stars, and the cause of their 
apparently large
contribution to the chemical evolution in $\omega$ Cen (Lloyd Evans 1983)
remains a mystery, although speculation as to the reason can be
found in Smith et al. (2000) and Cunha et al. (2002).  In this panel
no trend is found between [La/Fe] and [C/Fe], indicating that the
sources of La enrichment and C depletion are not related.  The more
La-poor giants in $\omega$ Cen again overlap perfectly with their
counterparts in M4. 

Possible correlations of $^{12}$C/$^{13}$C with abundances of three
elements either destroyed (O) or produced (Na and Al) by various
proton capture cycles are explored in Figure 7.  The top panel
plots the isotopic carbon ratios versus [O/Fe] for $\omega$ Cen,
M4, and M71.  In the regions of overlap between these three globular
clusters, there is excellent agreement between $^{12}$C/$^{13}$C
ratios and [O/Fe] abundance ratios.  When taken together, the three
cluster samples hint of a correlation between
$^{12}$C/$^{13}$C and [O/Fe].  This possible trend needs to be 
verified by future studies of larger samples, but it may
point towards a deep-mixing signature in the current
globular cluster giants.  The bottom two panels, with $^{12}$C/$^{13}$C
versus [Na/Fe] and [Al/Fe] show no hint of any trend.  This is
somewhat puzzling in light of the possible trend of the carbon isotopic
ratio with [O/Fe].  The deeper and more extensive mixing necessary to
enhance the [Na/Fe] and [Al/Fe] ratios, would be expected to drive 
the envelope C-isotope ratios to their CN-equilibrium value of 3.5:
this is not observed.  The lack of any correlation between
$^{12}$C/$^{13}$C and Na or Al points to the origins of some of these
abundance variations as arising in some other site, and not in the current
cluster giants, as concluded earlier by Briley et al. (1997). A larger 
sample of $^{12}$C, $^{13}$C, O, Na,
and Al abundances from globular cluster giants would be able to shed
more light on this issue. 

\section{Conclusions}

The $^{12}$C/$^{13}$C ratios in all of the $\omega$ Cen giants are
quite low, with a mean $\langle$$^{12}$C/$^{13}$C$\rangle$ = 4.3$\pm$0.4:
this indicates extensive and deep mixing in these low-mass, 
low-metallicity giants.  This result is in general agreement with
previous studies of globular cluster giants, as well as low-metallicity
field giants.  In particular for the $\omega$ Cen giants, all of which
have M$_{\rm V}$$\le$ -1.8, there is no meaurable change in
$^{12}$C/$^{13}$C over the metallicity range of [Fe/H]= -1.7 to
-0.7.  It is found that [$^{12}$C/Fe] correlates well with
[O/Fe] and anticorrelates with [Na/Fe], and that the overall trends
among these various ratios agrees very well between $\omega$ Cen and
other globular clusters.  The $^{12}$C/$^{13}$C ratios themselves
exhibit no measurable trends with [Na/Fe] or [Al/Fe], but may hint
at a positive correlation with [O/Fe].

\medskip
We wish to thank the staff of the Cerro Tololo Inter-American
Observatory, especially M.\ Fern\'andez and
H.\ Tirado, for their excellent assistance with 
the observations.  Partial support for this project came from the
National Science Foundation under grants AST-9157038
and INT-9215844 to The Ohio State University Research Foundation.
VVS acknowledges support from the National Science Foundation
through grant AST-9987374.

\clearpage

\figcaption[]{Color-magnitude diagram for proper motion
members of omega Cen from van Leeuwen et al. (2000), shown as
small symbols.  The photometry for the stars in the
current sample are shown as large points.
\label{figure1.ps}}

\figcaption[]{Sample spectra of two $\omega$ Cen giants.  Both of these
giants have similar effective temperatures and metallicities ([Fe/H]).
The prominent $^{12}$CO and $^{13}$CO bands are identified and
most of the structure observed as a function of wavelength is real
(being composed of many CO vibration-rotation lines).  The spectra are
normalized and shifted in relative flux so both can be viewed clearly.
Note that ROA 139 has weaker CO bands than ROA 155, indicating a lower
total carbon abundance in ROA 139, although both have quite low isotopic
ratios ($^{12}$C/$^{13}$C=4.5 for ROA 155 and 3.5 for ROA 139). 
\label{figure2.ps}}

\figcaption[]{Observed (solid dots) and synthetic (continuous curves)
spectra for the $\omega$ Cen giant ROA155 (T$_{\rm eff}$= 4200K,
log g= 0.8, $\xi$= 2.0 km s$^{-1}$, and [Fe/H]= -1.64).  The top panel
shows a region near the $^{12}$CO (3-1) bandhead, with synthetic spectra
calculated for three different $^{12}$C abundances (the best-fit
abundance and $\pm$0.1 dex).  The abundance notation A(x)=
log $\epsilon$(x)= log(N$_{\rm x}$/N$_{\rm H}$ + 12.0 is used.  The
bottom panel shows the $^{13}$CO (2-0) bandhead region with three
synthetic spectra computed with three different isotopic carbon abundance
ratios. 
\label{figure3.ps}}

\figcaption[]{A comparison of carbon-12 abundances (plotted as 
[$^{12}$C/Fe]) derived here using CO and derived by Norris \& Da Costa 
(1995b) using violet CH.  The solid line illustrates perfect agreement.
Errorbars are taken from the discussions in Norris \& Da Costa and
here.  The agreement between the two studies is good, with a mean
difference of (This Study -- Norris \& Da Costa)= +0.14$\pm$ 0.24 dex:
the scatter is that expected given the respective errors in each study.
\label{figure4.ps}}

\figcaption[]{Isotopic carbon abundance ratios are plotted as a function
of [Fe/H] (top panel) and absolute visual magnitude (bottom panel) for
the different samples of metal-poor giants.  There are no strong
trends with metallicity, although it should be noted that the scatter
in $^{12}$C/$^{13}$C in the more metal-rich globular clusters M71 and
47 Tuc is anticorrelated with the CN strengths (Briley et al. 1997).
The field sample also contains a number of giants with somewhat higher
isotopic ratios ($\sim$9--10) at lower metallicities; however, the
bottom panel reveals these field giants to be of somewhat lower
luminosities than most of the globular cluster giants analyzed.  The
bottom panel reveals a modest trend of decreasing $^{12}$C/$^{13}$C
as the giant's luminosity increases (as discussed by Keller et al. 2001):
the globular cluster giants fall on the same trend defined by the
field-giant population. 
\label{figure5.ps}}

\figcaption[]{Comparisons of four different abundance ratios versus
[$^{12}$C/Fe].  The top left panel shows $^{12}$C/$^{13}$C versus
[$^{12}$C/Fe] for $\omega$ Cen and a number of different samples of
low-metallicity giants.  No strong trends are evident, with all
giants mixed extensively.  The bottom left panel shows correlations
between [O/Fe] and [$^{12}$C/Fe] in three globular clusters: $\omega$
Cen, M4, and M71.  The respective trends for all three clusters
overlap and the correlation of depleted carbon-12 with oxygen is a
signature of material processed through the CNO cycles.  The top left
panel illustrates the anticorrelation between [Na/Fe] and [$^{12}$C/Fe]
in $\omega$ Cen and M4, again with substantial overlap in the two
clusters.  The decrease in carbon-12 is due to mixing, while the increase
in sodium is due, presumably, to the Na--Ne cycle.  The bottom right
panel explores the relation between lanthanum, an s-process element,
and $^{12}$C in $\omega$ Cen and M4.  No trends exist, indicating that
the mixing responsible for $^{12}$C depletion has no dependance on the
s-process production site. 
\label{figure6.ps}}

\figcaption[]{An investigation of possible trends of $^{12}$C/$^{13}$C
with O, Na, and Al.  In the top panel isotopic carbon ratios are
plotted versus [O/Fe] for $\omega$ Cen, M4, and M71.  The three
globular clusters together indicate a possible systematic decrease,
towards the CN-equilibrium value, in $^{12}$C/$^{13}$C as [O/Fe]
decreases: this might signify deep mixing in the observed giants. 
The middle and bottom panels show $^{12}$C/$^{13}$C versus [Na/Fe]
and [Al/Fe], respectively, with no measurable trends between these
abundance ratios. 
\label{figure7.ps}}

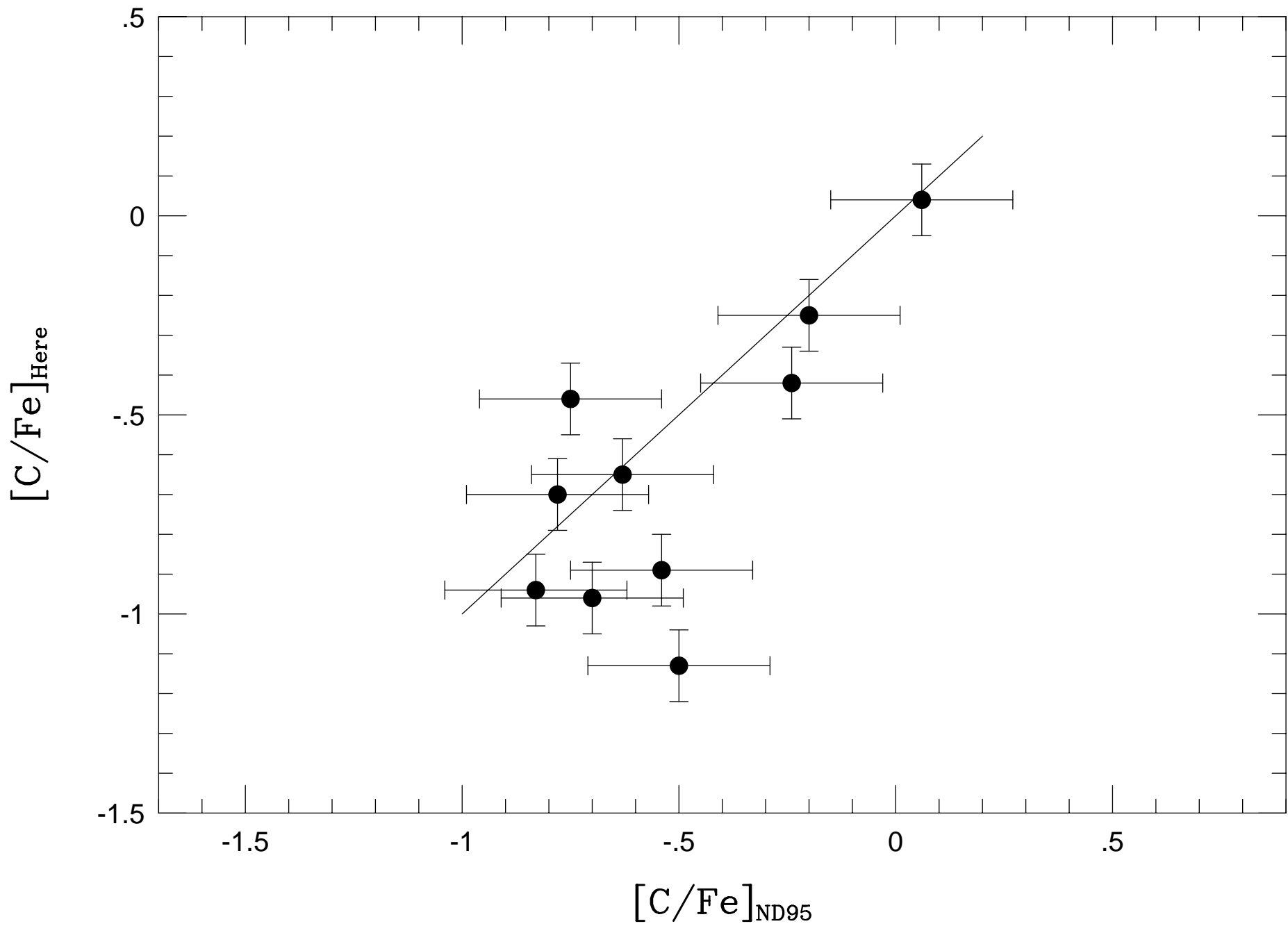
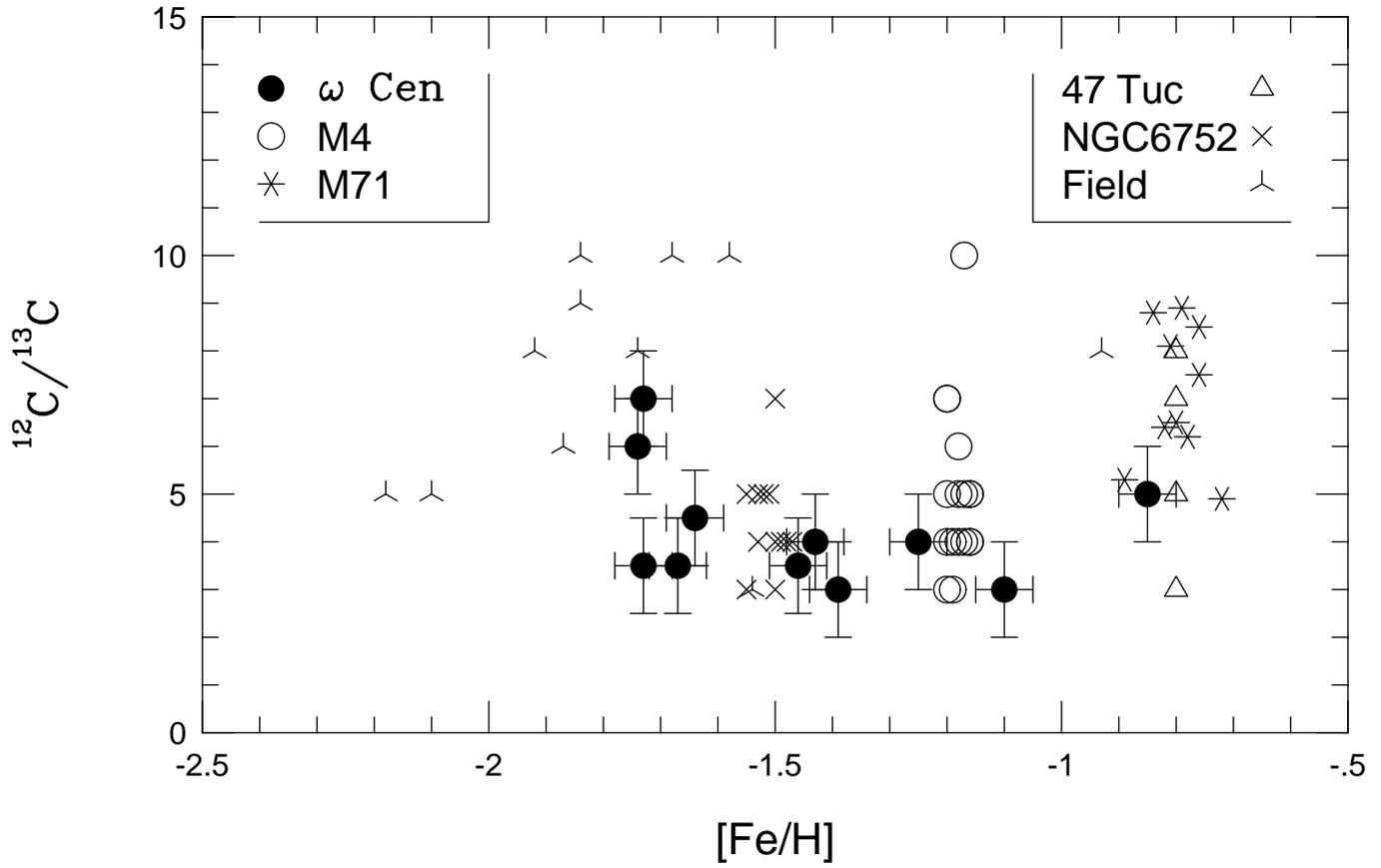
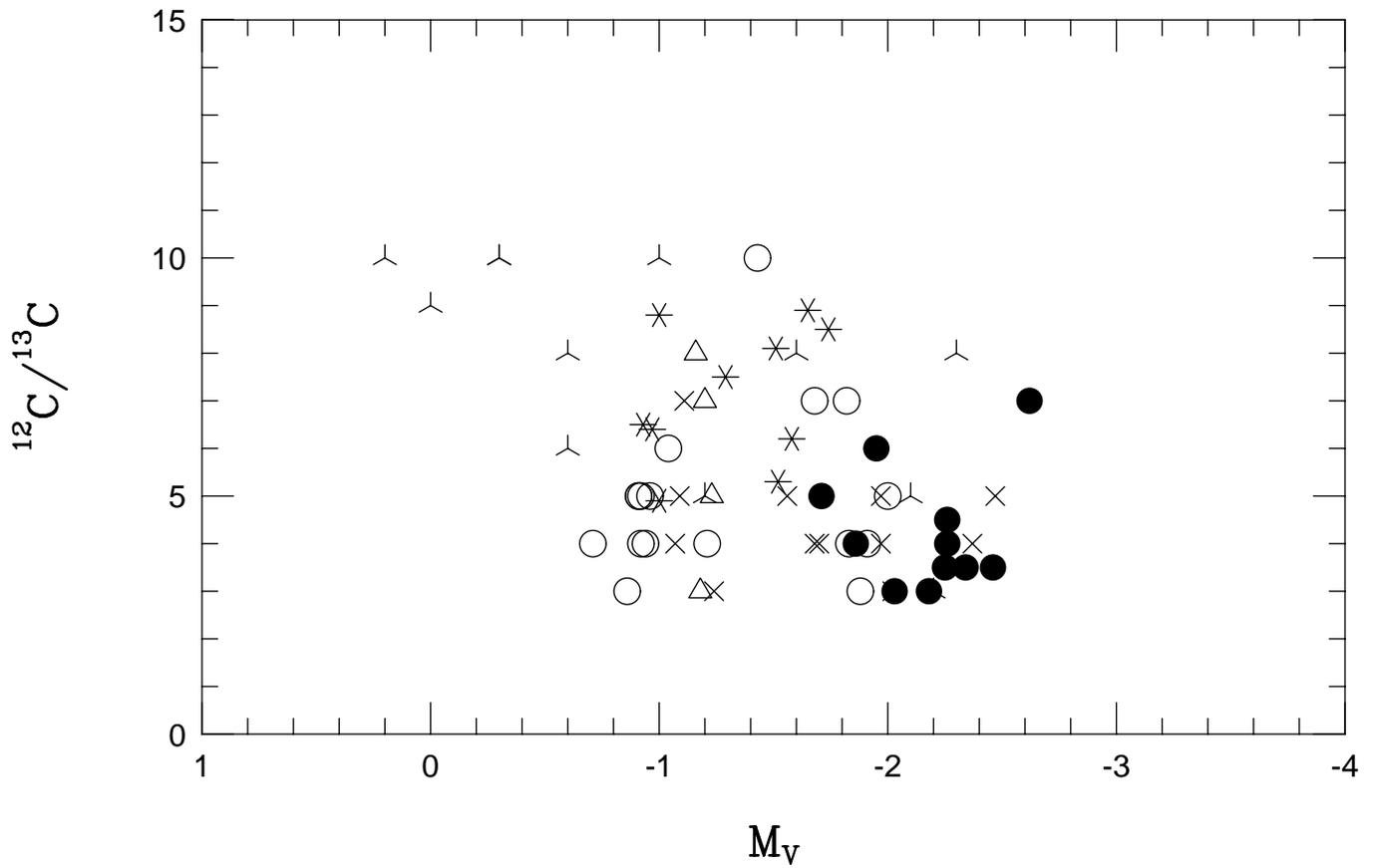
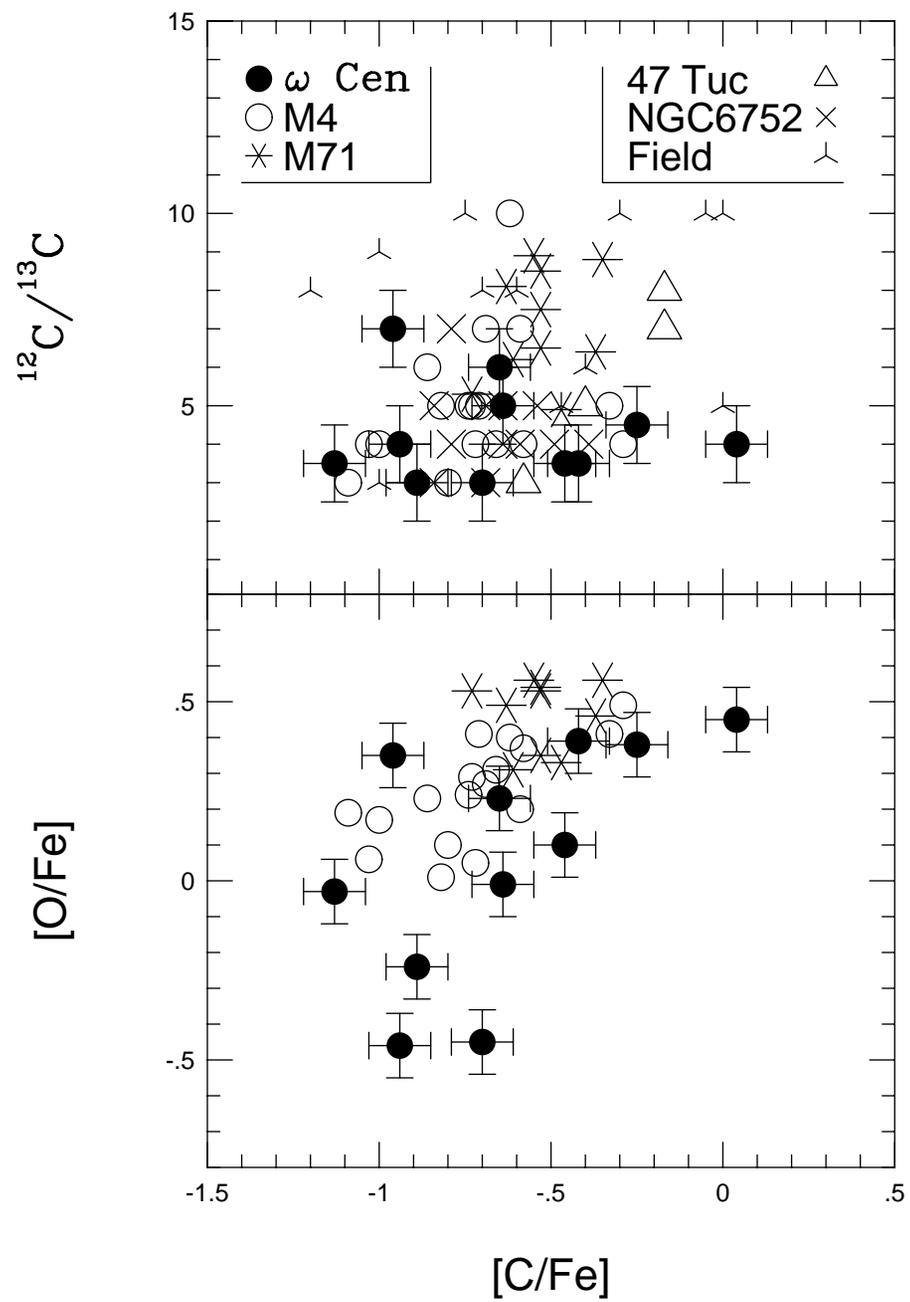
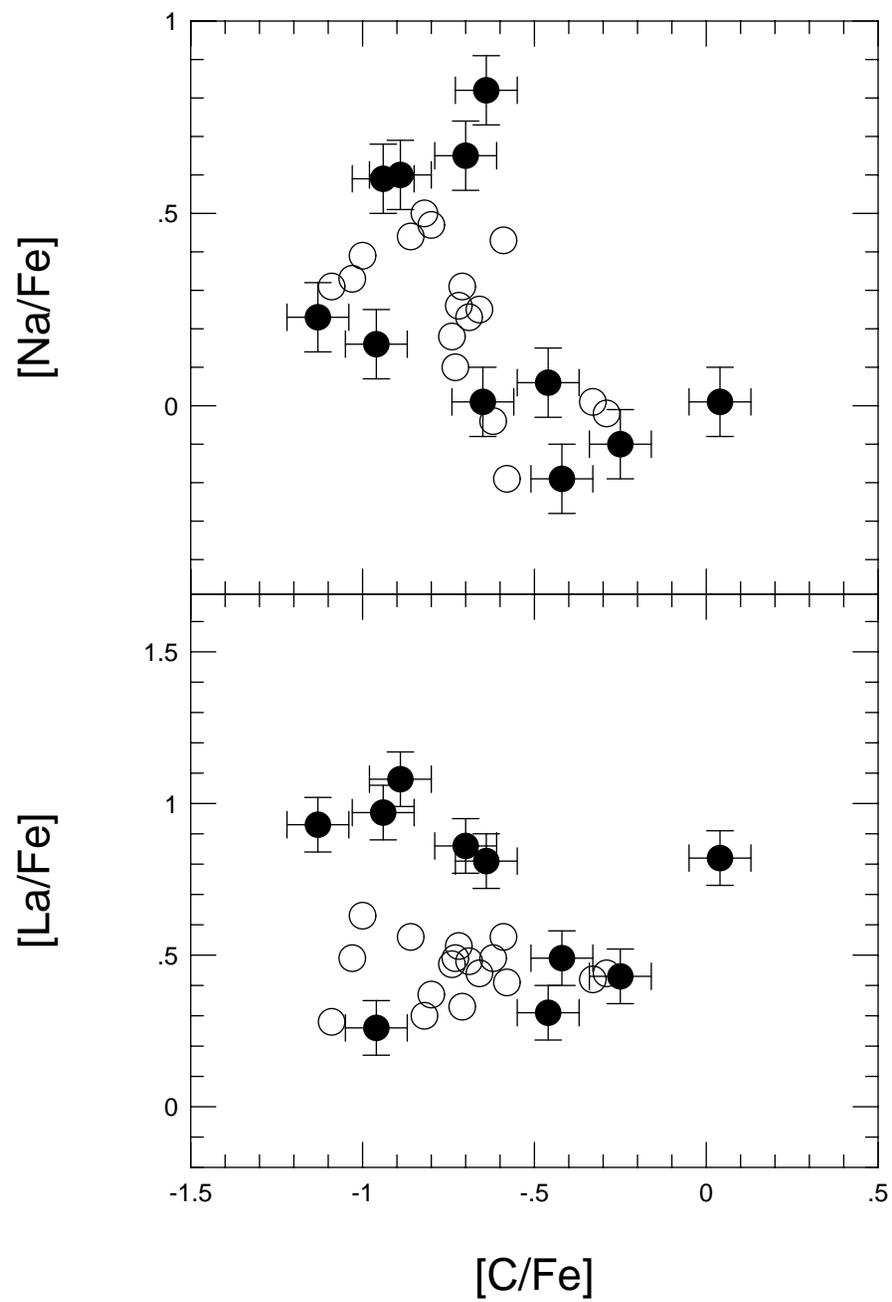
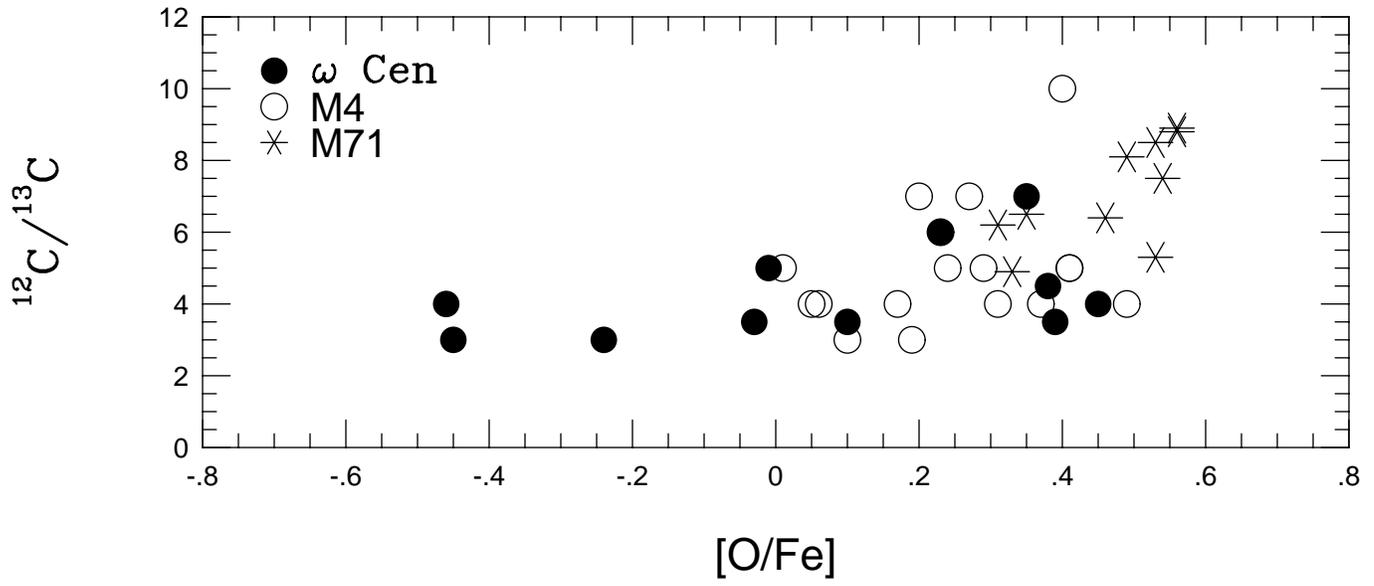
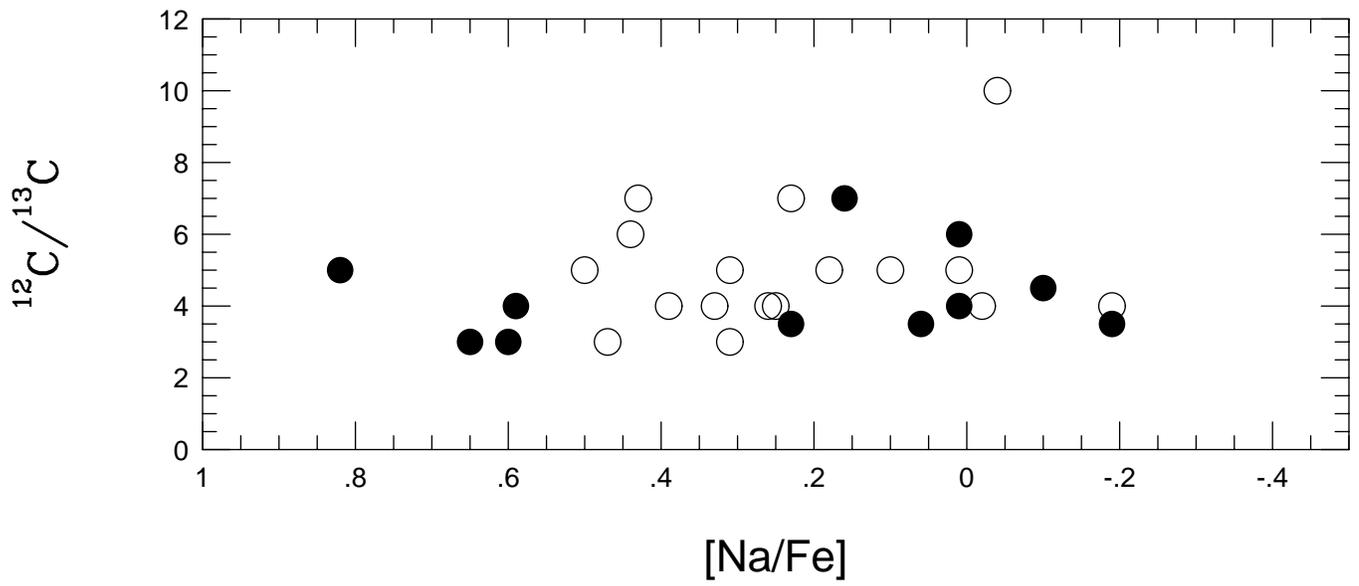
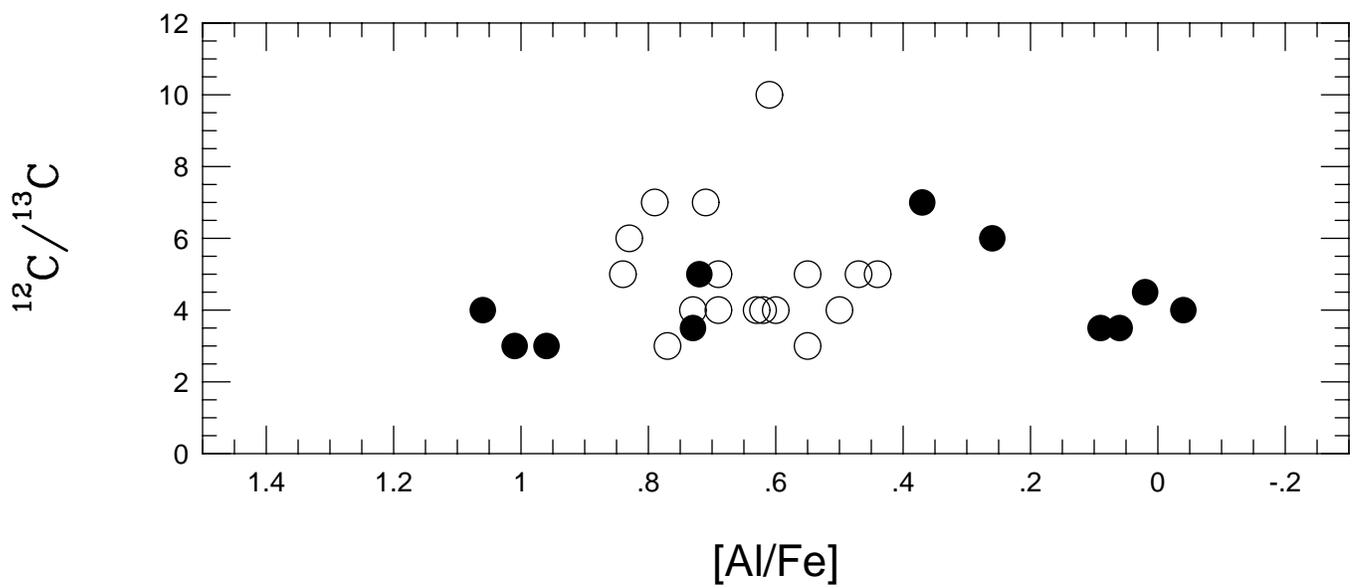
\end{document}